\documentclass[conference]{IEEEtran}

\usepackage{amsfonts,amsmath,amssymb}
\usepackage{dsfont}
\usepackage{cite}
\usepackage{citesort}
\usepackage{graphicx}
\usepackage{url}
\usepackage{caption}
\usepackage{bm}
\usepackage{bbding}
\usepackage{epstopdf}
\usepackage{algorithm}
\usepackage{algorithmic}
\usepackage{CJK}
\DeclareMathOperator*{\argmax}{argmax}

\newtheorem{theorem}{\bf{Theorem}}

\makeatletter
\def\ScaleIfNeeded{%
\ifdim\Gin@nat@width>\linewidth \linewidth \else \Gin@nat@width
\fi } \makeatother

\begin{document}

\title{Secure Transmission for Relay Wiretap Channels in the Presence of
Spatially Random Eavesdroppers}

\author{\IEEEauthorblockN{Chenxi Liu\IEEEauthorrefmark{2}, Nan
Yang\IEEEauthorrefmark{3}, Jinhong Yuan\IEEEauthorrefmark{2}, and
Robert Malaney\IEEEauthorrefmark{2}}
\IEEEauthorblockA{\IEEEauthorrefmark{2}School of Electrical
Engineering and Telecommunications, The University of New South
Wales, Sydney, Australia\\\IEEEauthorrefmark{3}Research School of
Engineering, Australian National University, Canberra, Australia}
Email: chenxi.liu@student.unsw.edu.au, nan.yang@anu.edu.au,
j.yuan@unsw.edu.au, r.malaney@unsw.edu.au}

\maketitle

\begin{abstract}
We propose a secure transmission scheme for a relay wiretap channel,
where a source communicates with a destination via a
decode-and-forward relay in the presence of spatially random-distributed eavesdroppers. We assume that the source is equipped
with multiple antennas, whereas the relay, the destination, and the
eavesdroppers are equipped with a single antenna each. In the
proposed scheme, in addition to information signals, the source transmits artificial noise signals
in order to confuse the
eavesdroppers. With the target of maximizing the secrecy throughput
of the relay wiretap channel, we derive a closed-form expression for
the transmission outage probability and an easy-to-compute
expression for the secrecy outage probability. Using these
expressions, we determine the optimal power allocation factor and
wiretap code rates that guarantee the maximum secrecy throughput,
while satisfying a secrecy outage probability constraint.
Furthermore, we examine the impact of source antenna number on the secrecy throughput, showing that adding extra transmit
antennas at the source brings about a significant increase in the
secrecy throughput.
\end{abstract}

\IEEEpeerreviewmaketitle

\section{Introduction}

Wireless communications are inherently insecure, due to the
broadcast nature of the medium, which makes security a pivotal
design issue in the implementation and operation of current and
future wireless networks. Compared to the traditional key-based
cryptographic techniques that are applied to upper layers,
physical layer security can enhance the secrecy of wireless
communications without using secret keys and complex
encryption/decryption algorithms, and thus has been recognized as an
alternative for cryptographic techniques. The key idea of physical
layer security is to exploit the randomness of wireless channels to
offer secure data transmissions \cite{Hong,Yang_Mag}. In early
studies, e.g., \cite{wyner}, the principle of physical layer
security was established in a single-input single-output wiretap
channel. Subsequently, physical layer security in
multi-input multi-output (MIMO) communication systems has been
intensively investigated
\cite{khisti,wornell,chenxi,chenxi2,chenxi3,Zhou10,Zhang13,nan,nan2,nan3,nan4,shihao},
due to the benefits of MIMO techniques such as high data rate and
high link reliability.

Most recently, physical layer security in large-scale wireless
networks, such as mobile ad hoc and sensor networks, has been
receiving considerable attention
\cite{xiangyun_2,wang_he,gio_1,gio_2}. A key property of large-scale
wireless networks is that the node locations in the network were
spatially randomly distributed. As such, stochastic geometry and
random geometric graphs are used for modeling the locations of
spatially random-distributed nodes. In \cite{xiangyun_2}, the
throughput of large-scale decentralized wireless networks with
physical layer security constraints was investigated. Considering
the path loss as the sole factor influencing the received
signal-to-noise ratio (SNR) at the receiver, \cite{wang_he} examined
the secrecy rate in cellular networks. In \cite{gio_1} the secrecy rate achieved by linear precoding
was analyzed in the broadcast channel with spatially
random external eavesdroppers, and in \cite{gio_2} the secrecy rate achieved by linear precoding in cellular networks was analyzed.


We note
that \cite{xiangyun_2,wang_he,gio_1,gio_2} only considered
point-to-point transmissions. This leaves physical layer security
with cooperative relays in large-scale networks as an open problem.
Since the relay channel efficiently improves the coverage and
reliability in wireless networks \cite{Laneman,Bassily}, it is of
practical significance to investigate the secrecy performance of
such channels. Particularly important in this context would be extensions of previous work on relay wiretap channels which focused on scenarios where the location of an eavesdropper(s) is fixed and known at
the source (e.g.,
\cite{dong,xiaoming_1,xiaoming_2,yulong,hanzhu,chenxi4}).


In this work we propose a secure transmission scheme for a relay
wiretap channel, where a source transmits to a destination via a
decode-and-forward (DF) relay in the presence of multiple spatially
random-distributed eavesdroppers. The source is
equipped with multiple antennas, while the relay, the destination,
and the eavesdropper are equipped with a single antenna each. In addition to information
signals, we assume that the source
transmit artificial noise signals in
order to confuse the eavesdroppers. Aiming at maximizing the secrecy throughput, while
satisfying a secrecy outage probability constraint, we determine
both the optimal power allocation factor between AN signals and
information signals and the optimal wiretap code rates.

Compared to current studies on physical layer security in relay
networks \cite{dong,xiaoming_1,xiaoming_2,yulong,hanzhu,chenxi4}, our contributions are threefold. First, we consider a more practical scenario where
the locations of the eavesdroppers are spatially randomly distributed
and not known at the source. Second, we propose a new secure
transmission scheme that maximizes the secrecy throughput in such a scenario. Third, we derive explicit expressions for the
transmission outage probability and the secrecy outage probability,
which enable us to analytically characterize the secrecy throughput
of the relay wiretap channel. We note that these expressions
are independent of realizations of the main channel and the
eavesdropper channel.


The rest of the paper is organized as follows. Section
\ref{sec:system_model} describes the system model and presents the proposed scheme. In Section
\ref{sec:analysis}, we analyze and optimize the secrecy performance
achieved by the proposed scheme. Numerical results and related
discussions are presented in Section \ref{sec:numerical}. Finally,
Section \ref{sec:conclusion} draws conclusions.

\section{System Model and Proposed Scheme}\label{sec:system_model}

\begin{figure}[!t]
\begin{center}{\includegraphics[width=0.9\columnwidth]{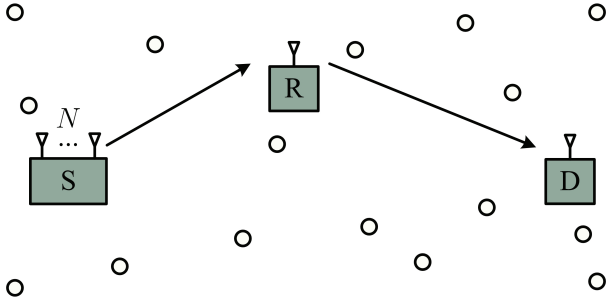}}
\caption{Illustration of a relay wiretap channel with spatially
random eavesdroppers.}\label{system_model}
\end{center}
\end{figure}

We consider a relay wiretap channel, as depicted in Fig.
\ref{system_model}, where a source (S) communicates with a
destination (D) with the aid of a relay (R) in the presence of
multiple spatially random eavesdroppers. We assume that the source
is equipped with $N$ antennas, while the destination, the relay, and
the eavesdroppers are equipped with a single antenna each. We also
assume that there is no direct link between the source and the
destination. We denote $d_{sr}$ and $d_{rd}$ as the distance between
the source and the relay and the distance between the relay and the
destination, respectively, and denote $\eta$ as the path loss
exponent. As in \cite{xiangyun_2,wang_he,gio_1,gio_2}, the eavesdroppers are modeled as a homogeneous Poisson
Point Process (PPP) $\Phi$ with density $\lambda$. This model is practical and representative for decentralized networks where each node is randomly distributed \cite{weber}. The source, the
relay, and the destination do not belong to $\Phi$.

\subsection{Transmission Scheme}

We now detail the proposed transmission scheme. The proposed scheme utilizes two time slots. In the first time
slot, the source transmits information signals together with AN
signals to the relay. In this time slot, the transmitted signals
from the source are overheard by the eavesdroppers. The role of the
AN signals is to confuse the eavesdroppers. In the second time slot, we assume that the source is silent. While it is clear that this will lead to a sub-optimal solution, such an assumption does allow for analytical tractability in a special case. Removing this assumption means more power can be added to the signal in the first time slot. Our other work \cite{chenxi5},  in which we consider multiple antennae at all nodes,  quantifies the advantage of adding source noise in the second time slot.
We also assume the relay transmits the received signals to the destination using
the DF protocol. In this time slot, the broadcast signals are also
overheard by the eavesdroppers. We assume that all the channels are
subject to independent and identically distributed Rayleigh fading.
We also assume a quasi-static block fading environment in which all
the channel coefficients remain the same within one time slot. We
denote $\mathbf{h}_{sr}$ as the $1\times N$ channel vector from the
source to the relay and denote $h_{rd}$ as the channel coefficient
from the relay to the destination.

We now express the transmitted and received signals in two time
slots separately. In the first time slot, the transmitted signal at
the source is given by
\begin{align}\label{x_s}
\mathbf{x}_{\text{S}} = \mathbf{W}\mathbf{t},
\end{align}
where $\mathbf{W}$ denotes the $N\times N$ beamforming matrix at the
source and $\mathbf{t}$ denotes the combination of the information
signal and the AN signal. To perform a such transmission, we first
design $\mathbf{W}$ as
\begin{align}\label{AN_beamforming}
\mathbf{W}=\begin{bmatrix}
\mathbf{w}_{\text{S}}&\mathbf{W}_{\text{AN}}
\end{bmatrix},
\end{align}
where $\mathbf{w}_{\text{S}}$ is used to transmit the information
signal and $\mathbf{W}_{\text{AN}}$ is used to transmit the AN
signal. The aim of $\mathbf{W}$ is to degrade the quality of the
received signals at the eavesdroppers. By transmitting AN signals
into the null space of $\mathbf{h}_{sr}$ through $\mathbf{W}$, the
source ensures that the quality of the received signals at the relay
is free from AN interference. In designing $\mathbf{W}$, we
choose $\mathbf{w}_{\text{S}}$ as the eigenvector corresponding to
the largest eigenvalue of $\mathbf{h}_{sr}^H\mathbf{h}_{sr}$. We
then choose $\mathbf{W}_{\text{AN}}$ as the remaining $N-1$
eigenvectors of $\mathbf{h}_{sr}^H\mathbf{h}_{sr}$. Such a design
ensures that $\mathbf{W}$ is a unitary matrix.
We then design $\mathbf{t}$ as
\begin{align}\label{t_s}
\mathbf{t}=\begin{bmatrix} t_{\text{S}}\\ \mathbf{t}_{\text{AN}}
\end{bmatrix},
\end{align}
where $t_{\text{S}}$ denotes the information signal and
$\mathbf{t}_{\text{AN}}$ is an $\left(N-1\right)\times 1$ vector of
the AN signal. We define $\beta$, $0<\beta\leq 1$, as the fraction
of the power allocated to the information signal. As such, we obtain
$\mathbb{E}\left[|t_{\text{S}}|^2\right]=\beta$ and
$\mathbb{E}\left[\mathbf{t}_{\text{AN}}\mathbf{t}_{\text{AN}}^H\right]=\frac{1-\beta}{N-1}\mathbf{I}_{N-1}$,
where $\mathbb{E}\left[\cdot\right]$ is expectation and
$\mathbf{I}_{m}$ is the $m\times{m}$ identity matrix. Moreover, we
confirm that
$\mathbb{E}\left[\mathbf{x}_{\text{S}}\mathbf{x}_{\text{S}}^H\right]=\mathbf{I}_N$.
Based on \eqref{x_s}, \eqref{AN_beamforming}, and \eqref{t_s}, the
received signal at the relay in the first time slot is expressed as
\begin{align}\label{signal_relay}
y_r =
\sqrt{P_sd_{sr}^{-\eta}}\mathbf{h}_{sr}\mathbf{w}_{\text{S}}t_{\text{S}}
+ n_r,
\end{align}
where $P_s$ denotes the transmit power at the source, and $n_r$ denotes the thermal noise at the relay, which is assumed to be a zero
mean complex Gaussian random variable with variance $\sigma_r^2$,
i.e., $n_r\sim\mathcal{CN}\left(0,\sigma_r^2\right)$.

We next express the received signal at a typical eavesdropper
located at $i$, $i\in\Phi$, in the first time slot as
\begin{align}
\label{signal_i} y_i^{(1)} =
\sqrt{P_sd_{si}^{-\eta}}\mathbf{h}_{si}\mathbf{w}_{\text{S}}t_{\text{S}}
+
\sqrt{P_sd_{si}^{-\eta}}\mathbf{h}_{si}\mathbf{W}_{\text{AN}}\mathbf{t}_{\text{AN}}
+n_{i1},
\end{align}
where $\mathbf{h}_{si}$ denotes the $1\times N$ channel vector from
the source to the typical eavesdropper located at $i$, $d_{si}$
denotes the distance between the source and the typical eavesdropper
located at $i$, and $n_{i1}$ denotes the thermal noise at the
typical eavesdropper located at $i$, which is assumed to be a zero mean complex
Gaussian random variable with variance $\sigma_{i1}^2$, i.e.,
$n_{i1}\sim\mathcal{CN}\left(0,\sigma_{i1}^2\right)$.

In the second time slot, the relay adopts the DF protocol to
forward signals to the destination.
Specifically, the relay first decodes the transmitted signals from
the source and then broadcasts them after re-encoding. Therefore, we
express the received signal at the destination as
\begin{align}\label{signal_d}
y_d=\sqrt{P_rd_{rd}^{-\eta}}h_{rd}x_{\text{R}}+n_d,
\end{align}
where $P_r$ denotes the transmit power at the relay, $h_{rd}$ denotes
the channel coefficient from the relay to the destination, $x_{\text{R}}$ denotes the transmitted signal of the relay
with $\mathbb{E}\left[\|x_{\text{R}}\|^2\right]=1$, and $n_d$
denotes the thermal noise at the destination, which is assumed to be a zero mean
complex random variable with variance $\sigma_d^2$, i.e.,
$n_d\sim\mathcal{CN}\left(0,\sigma_d^2\right)$.

We next express the received signal in the second time slot at the typical eavesdropper
located at $i$ is expressed as
\begin{align}\label{signal_i2}
y_i^{(2)}=\sqrt{P_rd_{ri}^{-\eta}}h_{ri}x_{\text{R}}+n_{i2},
\end{align}
where $h_{ri}$ denotes the channel coefficient from the relay to the
typical eavesdropper located at $i$, $d_{ri}$ denotes the distance
between the relay and the typical eavesdropper located at $i$, and
$n_{i2}$ denotes the thermal noise at the the typical eavesdropper
located at $i$, which is assumed to be a zero mean complex random variable with
variance $\sigma_{i2}^2$, i.e.,
$n_{i2}\sim\mathcal{CN}\left(0,\sigma_{i2}^2\right)$.

\subsection{Received SNRs}

We express the
instantaneous SNR at the destination as
\begin{align}
\label{snr_d} \Gamma_D = \min\left\{\gamma_{sr},\gamma_{rd}\right\},
\end{align}
where
$\gamma_{sr}=\frac{\beta{}P_s}{d_{sr}^{\eta}\sigma_r^{2}}\|\mathbf{h}_{sr}\|^2$
and
$\gamma_{rd}=\frac{P_r}{d_{rd}^{\eta}\sigma_d^{2}}\|{h}_{rd}\|^2$.
For the instantaneous SNR at the eavesdroppers, we assume that the
eavesdroppers are non-colluding, which means that each
eavesdropper decodes her own received signals from the source and
the relay, without cooperating with other eavesdroppers. Thus, we express the instantaneous SNR at the
eavesdroppers as\footnote{We assume that the source and the relay use different wiretap codes with
different codebooks. As such, the transmitted signals from the source and relay cannot be jointly processed by any  eavesdropper (nor by any combination of eavesdroppers).}
\begin{align}\label{snr_e}
\Gamma_E=\max_{i\in\Phi}\left\{\max\left\{\gamma_{si},\gamma_{ri}\right\}\right\},
\end{align}
where
\begin{align}\label{snr_si}
\gamma_{si}=\frac{\beta{}P_sd_{si}^{-\eta}\|\mathbf{h}_{si}\mathbf{w}_{\text{S}}\|^2}
{\frac{1-\beta}{N-1}P_sd_{si}^{-\eta}\mathbf{h}_{si}\mathbf{W}_{\text{AN}}\mathbf{W}_{\text{AN}}^H\mathbf{h}_{si}^H+\sigma_{i1}^2}.
\end{align}
and
$\gamma_{ri}=\frac{P_r}{d_{ri}^{\eta}\sigma_{i2}^2}\|h_{ri}\|^2$.

\subsection{Problem Formulation}

In order to evaluate and optimize the secrecy performance achieved
by our proposed scheme, we apply the
performance metric proposed in \cite{Zhou10}, which is given by
\begin{align}\label{throughput}
T_s=\frac{1}{2}\left(R_b-R_e\right)\left(1-P_{to}\right),
\end{align}
where the factor $1/2$ is due to the two time slots used in the transmission, $\left(R_b,R_e\right)$ denotes a parameter pair of the wiretap
code used by the source, $R_b$ denotes the transmission rate of the
wiretap code, $R_e$ denotes the cost of preventing the transmitted
wiretap code from eavesdropping, and $P_{to}$ denotes the
transmission outage probability ($P_{to}$ is defined as the probability
that the instantaneous SNR at the destination is less than
$\tau_b = 2^{R_b}-1$). Mathematically, $P_{to}$ is formulated as
\begin{align}\label{p_to}
P_{to}={\Pr}\left(\Gamma_{D}\leq\tau_b\right).
\end{align}
Henceforth, we refer to $T_s$ as the secrecy throughput.

The goal of this work is to maximize the secrecy throughput of the relay
wiretap channel with spatially random eavesdroppers under secrecy
constraints. To achieve this goal, we formulate the design problem
as
\begin{align}\label{prob_form}
\max_{R_b,R_e,\beta}& ~~T_s,\notag\\
s.t.~~P_{so}\leq \varphi, 0\leq R_e&\leq R_b, 0<\beta\leq1,
\end{align}
where $P_{so}$ denotes the secrecy outage probability. $P_{so}$ is
defined as the probability that $\Gamma_{E}$ is larger than $\tau_e
= 2^{R_e}-1$. Mathematically, $P_{so}$ is formulated as
\begin{align}\label{p_sec}
P_{so}={\Pr}\left(\Gamma_{E}>\tau_e\right).
\end{align}

\section{Analysis and Optimization of Secrecy Performance}\label{sec:analysis}

In this section, we first analyze the secrecy performance by
deriving explicit expressions for the transmission probability and
the secrecy outage probability, respectively. Based on these
results we determine the optimal parameters, e.g., $R_b$, $R_e$,
and $\beta$, that achieve the optimal secrecy performance of the
relay wiretap channel. Notably, the determined optimal secrecy
performance is independent of realizations of the main channel and
the eavesdroppers' channels.


\subsection{Transmission Outage Probability}

In this subsection we focus on the transmission outage probability,
$P_{to}$. We first obtain the cumulative distribution functions
(CDFs) of $\gamma_{sr}$ and $\gamma_{rd}$ as
\begin{align}\label{cdf_1}
F_{\gamma_{sr}}\left(\gamma\right)=1-\exp\left(-\frac{\gamma}{\beta\overline{\gamma}_{sr}}\right)
\sum_{n=0}^{N-1}\frac{1}{n!}\left(\frac{\gamma}{\beta\overline{\gamma}_{sr}}\right)^n
\end{align}
with $\overline{\gamma}_{sr}=P_sd_{sr}^{-\eta}\sigma_r^{-2}$, and
\begin{align}\label{cdf_2}
F_{\gamma_{rd}}\left(\gamma\right)=1-\exp\left(-\frac{\gamma}{\overline{\gamma}_{rd}}\right)
\end{align}
with $\overline{\gamma}_{rd}=P_rd_{rd}^{-\eta}\sigma_d^{-2}$.
Based on \eqref{snr_d}, \eqref{cdf_1}, and \eqref{cdf_2}, we
re-express the transmission outage probability in \eqref{p_to} as
\begin{align}\label{p_tr_2}
P_{to}=&{\Pr}\left(\min\left\{\gamma_{sr},\gamma_{rd}\right\}\leq\tau_b\right)\notag\\
=&1-\left(1-F_{\gamma_{sr}}\left(\tau_b\right)\right)\left(1-F_{\gamma_{rd}}\left(\tau_b\right)\right)\notag\\
=&1-\exp\left(-\left(\frac{1}{\beta\overline{\gamma}_{sr}}+\frac{1}{\overline{\gamma}_{rd}}\right)\tau_b\right)
\sum_{n=0}^{N-1}\frac{1}{n!}\left(\frac{\tau_b}{\beta\overline{\gamma}_{sr}}\right)^n.
\end{align}

\subsection{Secrecy Outage Probability}

We now focus on the secrecy outage probability, $P_{so}$. To derive
$P_{so}$, we first express the CDFs of $\gamma_{si}$ and
$\gamma_{ri}$ as
\begin{align}\label{cdf_3}
F_{\gamma_{si}}\left(\gamma\right)=1-\left(1+\frac{\left(1-\beta\right)\gamma}
{\beta\left(N-1\right)}\right)^{-\left(N-1\right)}\exp\left(-\frac{\gamma}{\beta\overline{\gamma}_{si}}\right)
\end{align}
with $\overline{\gamma}_{si}=P_sd_{si}^{-\eta}\sigma_{i1}^{-2}$, and
\begin{align}\label{cdf_4}
F_{\gamma_{ri}}\left(\gamma\right)=1-\exp\left(-\frac{\gamma}{\overline{\gamma}_{ri}}\right)
\end{align}
with $\overline{\gamma}_{ri} = P_rd_{ri}^{-\eta}\sigma_{i2}^{-2}$.
Based on \eqref{snr_e}, \eqref{cdf_3} and \eqref{cdf_4}, we
present the secrecy outage probability in the following theorem.
\begin{theorem}\label{t1}
The secrecy outage of the relay wiretap channel is derived as
\begin{align}\label{t1_result}
P_{so}=1-\exp\left(-2\lambda\left(\mathcal{J}_1 + \mathcal{J}_2 - \mathcal{J}_3\right)\right),
\end{align}
where
\begin{align}\label{t1_result_2}
\mathcal{J}_{1}=&\frac{\pi}{\eta}\left(\frac{\beta
P_s}{\tau_e\sigma_{i1}^2}\right)^{\frac{2}{\eta}}
\left(1+\frac{\left(1-\beta\right)\tau_e}{\beta\left(N-1\right)}\right)^{-\left(N-1\right)}\Gamma\left(\frac{2}{\eta}\right),
\end{align}
\begin{align}\label{t1_result_3}
\mathcal{J}_{2}=\int_{0}^{\infty}\int_{0}^{\pi}d_{si}\exp\left(-\psi\left(\theta\right)\right)dd_{si}d\theta,
\end{align}
\begin{align}\label{t1_result_4}
\mathcal{J}_{3}=&\left(1+\frac{\left(1-\beta\right)\tau_e}{\beta\left(N-1\right)}\right)^{-\left(N-1\right)}\notag\\
&\times\int_{0}^{\infty}\int_{0}^{\pi}d_{si}\exp\left(-\frac{\tau_e\sigma_{i1}^2}{\beta{}P_s}d_{si}^{\eta}\right)
\exp\left(-\psi\left(\theta\right)\right)dd_{si}d\theta,
\end{align}
and
$\psi\left(\theta\right)=\frac{\tau_e\sigma_{i2}^2}{P_r}\left(d_{sr}^2+d_{si}^2-2d_{sr}d_{si}\cos\theta\right)^{\frac{\eta}{2}}$.
\begin{IEEEproof}
See Appendix \ref{App_t1}.
\end{IEEEproof}
\end{theorem}

We find that \eqref{t1_result} provides an easy-to-compute
expression for the secrecy outage probability. Despite that
$\mathcal{J}_2$ and $\mathcal{J}_3$ for general $\eta$ can not be
obtained in closed-form, they can be easily
evaluated since only a double integral is
involved in $\mathcal{J}_2$ and $\mathcal{J}_3$.

We next present simplified closed-form expressions for the special
case where $\eta=2$. For this special case, we first simplify
$\mathcal{J}_2$, which yields a closed-form expression given by
\eqref{J_2_fur} (next page),
\begin{figure*}[ht]
\begin{align}\label{J_2_fur}
\mathcal{J}_2 = & \exp\left(-\frac{\tau_e\sigma_{i2}^2}{P_r}d_{sr}^2\right)\int_0^{\infty}d_{si}\exp\left(-\frac{\tau_e\sigma_{i2}^2}{P_r}d_{si}^2\right)
\int_{0}^{\pi}\exp\left(\frac{2\tau_e\sigma_{i2}^2}{P_r}d_{sr}d_{si}\cos\theta\right)d\theta dd_{si}\notag\\
\overset{(a)}{=}&\pi\exp\left(-\frac{\tau_e\sigma_{i2}^2}{P_r}d_{sr}^2\right)\int_0^{\infty}d_{si}\exp\left(-\frac{\tau_e\sigma_{i2}^2}{P_r}d_{si}^2\right)
I_{0}\left(\frac{2\tau_e\sigma_{i2}^2}{P_r}d_{sr}d_{si}\right)dd_{si}\notag\\
\overset{(b)}{=}&\pi\exp\left(-\frac{\tau_e\sigma_{i2}^2}{P_r}d_{sr}^2\right)\int_{0}^{\infty}d_{si}\exp\left(-\frac{\tau_e\sigma_{i2}^2}{P_r}d_{si}^2\right)
\sum_{k=0}^{\infty}\frac{1}{\left(k!\right)^2}\left(\frac{\tau_e\sigma_{i2}^2}{P_r}d_{sr}d_{si}\right)^{2k}dd_{si}\notag\\
\overset{(c)}{=}&\frac{\pi P_r}{2\tau_e\sigma_{i2}^2}\exp\left(-\frac{\tau_e\sigma_{i2}^2}{P_r}d_{sr}^2\right)
\sum_{k=0}^{\infty}\frac{1}{\left(k!\right)^2}\left(\frac{\tau_e\sigma_{i2}^2}{P_r}d_{sr}^2\right)^k\Gamma\left(k+1\right)\notag\\
\overset{(d)}{=}&\frac{\pi P_r}{2\tau_e\sigma_{i2}^2}.
\end{align}
\hrule
\end{figure*}
where $(a)$ follows by applying the expression of the zero-order
modified Bessel function $I_0(z)$, $(b)$
follows by applying the series representation of $I_0(z)$, $(c)$ follows by  using $t=\frac{\tau_e\sigma_{i2}^2}{P_r}d_{si}^{2}$ and noting that
\begin{align}
\int_{0}^{\infty}t^{k}\exp\left(-t\right)dt=\Gamma\left(k+1\right),
\end{align}
and $(d)$ follows by applying the series expression representation
of the exponential function. Similarly,
for the special case where $\eta=2$, we simplify $\mathcal{J}_3$,
which yields a closed-form expression given by \eqref{J_3_fur} (next page). The simplified expressions in \eqref{J_2_fur} and
\eqref{J_3_fur} offer us a computationally efficient way to calculate the secrecy outage probability for the special case where $\eta=2$.
\begin{figure*}[ht]
\begin{align}\label{J_3_fur}
\mathcal{J}_3 = & \left(1+\frac{\left(1-\beta\right)\tau_e}{\beta\left(N-1\right)}\right)^{-\left(N-1\right)}\!\!\exp\!\left(\!-\frac{\tau_e\sigma_{i2}^2}{P_r}d_{sr}^2\!\right)\!
\int_{0}^{\infty}\!d_{si}\exp\!\left(\!-\left(\frac{\tau_e\sigma_{i1}^2}{\beta P_s}+\frac{\tau_e\sigma_{i2}^2}{P_r}\right)d_{si}^2\!\right)\!
\int_{0}^{\pi}\exp\left(\frac{2\tau_e\sigma_{i2}^2}{P_r}d_{sr}d_{si}\cos\theta\right)d\theta d d_{si}\notag\\
=&\frac{\pi\beta P_s P_r}{2\tau_e\left(\beta P_s\sigma_{i2}^2 + P_r\sigma_{i1}^2\right)}
\exp\left(-\frac{\tau_e\sigma_{i1}^2\sigma_{i2}^2}{\beta P_s\sigma_{i2}^2+P_r\sigma_{i1}^2}d_{sr}^2\right)
\left(1+\frac{\left(1-\beta\right)\tau_e}{\beta\left(N-1\right)}\right)^{-\left(N-1\right)}.
\end{align}
\hrule
\end{figure*}

\subsection{Throughput Optimization}

In this subsection we determine the optimal parameters that
maximize the secrecy throughput, $T_s$, for general $\eta$. Specifically, we first determine the optimal wiretap code rates
pair, $\left(R_b^{\ast},R_e^{\ast}\right)$, that maximizes the
secrecy throughput for a given power allocation factor $\beta$. We
then determine the joint optimal power allocation factor and wiretap
code rates,
$\left(\beta^{\ast\circ},R_b^{\ast\circ},R_e^{\ast\circ}\right)$,
that maximizes $T_s$.

\subsubsection{Optimal wiretap code rates pair for a given $\beta$}

The optimal wiretap code rates pair,
$\left(R_b^{\ast},R_e^{\ast}\right)$, that maximizes $T_s$ for a given $\beta$ is determined as
\begin{align}\label{prob_form_2}
\left(R_b^{\ast},R_e^{\ast}\right) = \argmax~ T_s,\notag\\
s.t.~~ P_{so}\leq\varphi, 0\leq R_e\leq R_b.
\end{align}
Taking the first-order derivative of $P_{so}$ with respect to $R_e$,
we confirm that $\partial P_{so}/\partial R_e<0$, which states
that $P_{so}$ monotonically decreases as $R_e$ increases. As such,
the value of $R_e^{\ast}$ satisfying \eqref{prob_form_2} is the
value of $R_e^{\ast}$ that satisfies the secrecy outage probability
constraint, i.e., $P_{so}\left(R_e^{\ast}\right)=\varphi$. We
then confirm that $\partial T_s/\partial R_b$ is first positive then
negative as $R_b$ increases. This demonstrates that the value of
$R_b^{\ast}$ satisfying \eqref{prob_form_2} is unique. Although a
closed-form solution for $\left(R_b^{\ast},R_e^{\ast}\right)$ is
mathematically intractable, we are able to determine the values of
$\left(R_b^{\ast},R_e^{\ast}\right)$ numerically. Accordingly, the
maximal secrecy throughput based on the values of
$\left(R_b^{\ast},R_e^{\ast}\right)$ for a given $\beta$ is defined
as $T_s^{\ast}$.

\subsubsection{Joint optimization of $\beta$, $R_b$, and $R_e$}

The joint optimal power allocation factor and wiretap code rates
which maximizes $T_s$ in \eqref{throughput},
$\left(\beta^{\ast\circ},R_b^{\ast\circ},R_e^{\ast\circ}\right)$, is
determined as
\begin{align}\label{prob_form_3}
\left(\beta^{\ast\circ},R_b^{\ast\circ},R_e^{\ast\circ}\right)& = \argmax~ T_s,\notag\\
s.t.~~ P_{so}\leq\varphi, ~0\leq R_e&\leq R_b, ~0<\beta\leq1.
\end{align}
Using \eqref{p_tr_2} and \eqref{t1_result}, we are able to solve
\eqref{prob_form_3} numerically. Specifically, we first determine
the value of $\left(R_b^{\ast},R_e^{\ast}\right)$ using
\eqref{prob_form_2} for each value of $\beta$. This leads to the
secrecy throughput with $\left(R_b^{\ast},R_e^{\ast}\right)$,
denoted by
$T_s^{\ast}=\frac{1}{2}\left(R_b^{\ast}-R_e^{\ast}\right)\left(1-P_{to}\right)$.
We then determine the value of $\beta$ that maximizes $T_s^{\ast}$,
denoted by $\beta^{\ast\circ}$. Accordingly, the value of
$\left(R_b^{\ast},R_e^{\ast}\right)$ associated with
$\beta^{\ast\circ}$ is determined as
$\left(R_b^{\ast\circ},R_e^{\ast\circ}\right)$. Finally, the maximal
secrecy throughput based on the values of $\beta^{\ast\circ}$,
$R_b^{\ast\circ}$, and $R_e^{\ast\circ}$ is defined as
$T_s^{\ast\circ}$.

\section{Numerical Results}\label{sec:numerical}

In this section we present numerical results to validate our
analysis of the outage probabilities and examine the benefits of the
proposed scheme. For illustrative purpose, throughout this section we concentrate on the
practical example of a highly shadowed urban area with $\eta = 4$. In addition, we adopt $\lambda d_{sr}^2 = \lambda d_{rd}^2 =
1$.

\begin{figure}[t!]
\begin{center}{\includegraphics[height=2.3in,width=3.0in]{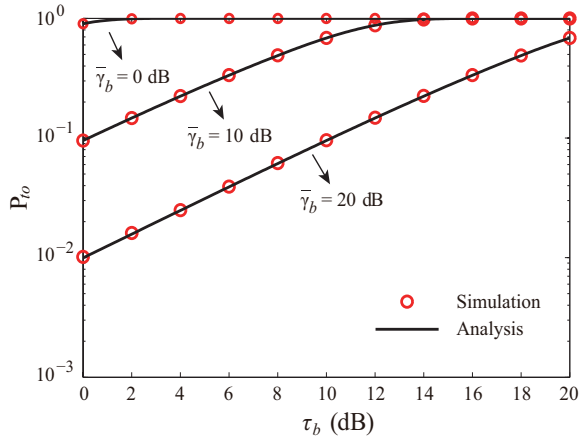}}
\caption{$P_{to}$ versus $\tau_b$ for different values of
$\overline{\gamma}_b$ with $\eta=4$, $N=4$, and $\beta =
0.5$.}\label{fig_side_a}
\end{center}
\end{figure}

We first verify the accuracy of the transmission outage probability
and the secrecy outage probability using Monte Carlo simulations. In
Fig. \ref{fig_side_a}, we plot $P_{to}$ versus $\tau_b$ for
different values of $\overline{\gamma}_b$ with $N=4$ and $\beta=0.5$. In this
figure, we consider
$\overline{\gamma}_{sr}=\overline{\gamma}_{rd}=\overline{\gamma}_b$.
We first see that the analytical curves, generated from
\eqref{p_tr_2}, precisely match the simulation points, which
demonstrates the correctness of our expression for $P_{to}$ in
\eqref{p_tr_2}. Second, we see that $P_{to}$ increases monotonically
as $\tau_b$ increases for a given $\overline{\gamma}_b$, which
implies that the transmission outage probability increases when the
transmission rate of the wiretap code increases. We further see that
$P_{to}$ decreases as $\overline{\gamma}_b$ increases for a given
$\tau_b$. This reveals that the transmission outage probability
reduces when the source and the relay use more power to transmit
under a fixed $\tau_b$.

\begin{figure}[t!]
\begin{center}{\includegraphics[height=2.3in,width=3.0in]{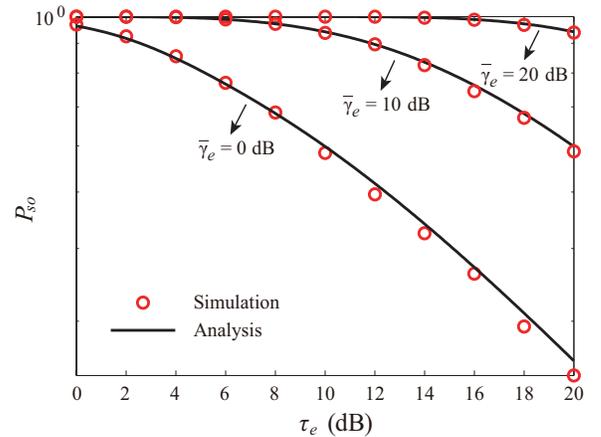}}
\caption{$P_{so}$ versus $\tau_e$ for different values of
$\overline{\gamma}_e$ with $\eta=4$, $N=4$, and $\beta =
0.5$.}\label{fig_side_b}
\end{center}
\end{figure}

In Fig. \ref{fig_side_b}, we plot $P_{so}$ versus $\tau_e$ for
different values of $\overline{\gamma}_e$ with $N=4$ and $\beta = 0.5$. In
this figure we consider
$\overline{\gamma}_{sr}{\sigma_r^2}/{\sigma_{i1}^2}=\overline{\gamma}_{rd}{\sigma_d^2}/{\sigma_{i2}^2}=\overline{\gamma}_e$.
We see an excellent match between the analytical curves
generated from \eqref{t1_result} and the simulation points,
demonstrating the correctness of our expression for $P_{so}$ in
\eqref{t1_result}. We then see that $P_{so}$ decreases monotonically
as $\tau_e$ increases for a given $\overline{\gamma}_e$, which
shows that the secrecy outage probability decreases when the
redundancy rate of the wiretap code increases. We further observe
that $P_{so}$ increases as $\overline{\gamma}_e$ increases. This is
due to the fact the eavesdroppers receive signals from {\it both} the
source and the relay. It follows that increasing the transmit power
at the source and the relay leads to an improved received SNR at the
eavesdroppers.

\begin{figure}[t!]
\begin{center}{\includegraphics[height=2.3in,width=3.0in]{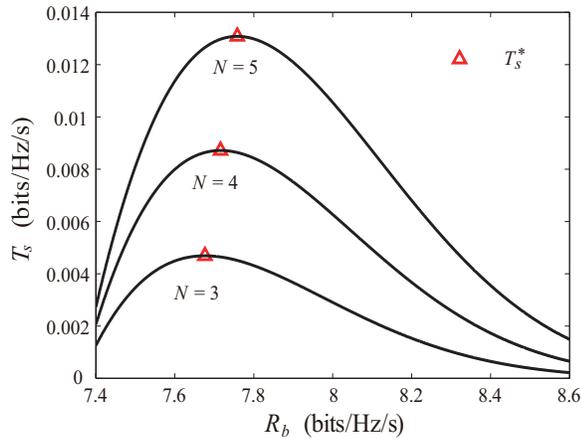}}
\caption{$T_s$ versus $R_b$ for different values of $N$ with
$\eta=4$, $\beta = 0.5$, $\varphi = 0.4$, $\overline{\gamma}_b = 20$
dB, and $\overline{\gamma}_b/\overline{\gamma}_e =
20$.}\label{fig_side_c}
\end{center}
\end{figure}

We now examine the impact of the system parameters $R_b$ and $\beta$ on
the secrecy throughput. In Fig. \ref{fig_side_c}, we plot $T_s$
versus $R_b$ for different values of $N$ with a fixed $\beta$ and the optimal $R_e$. We first observe that there exists a
unique $R_b^{\ast}$ that maximizes $T_s$ for a given $\beta$. We
also observe that the maximal $T_s$ for a given $\beta$, i.e.,
$T_s^{\ast}$, increases as $N$ increases. This shows that adding
extra transmit antennas at the source significantly enhances the
secrecy performance of the relay wiretap channel.

\begin{figure}[t!]
\begin{center}{\includegraphics[height=2.3in,width=3.0in]{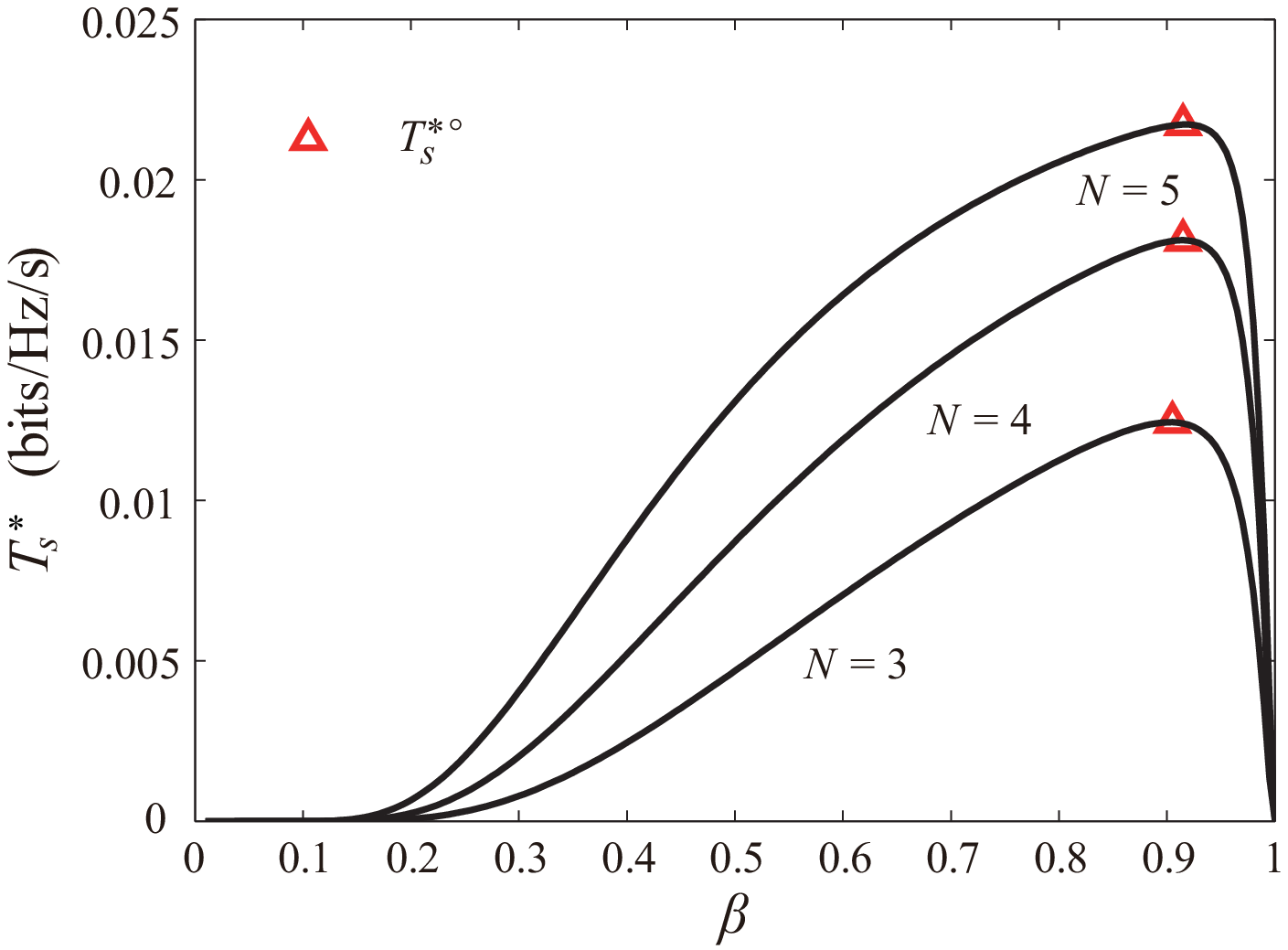}}
\caption{$T_s^{\ast}$ versus $\beta$ for different values of $N$
with $\eta=4$, $\varphi = 0.4$, $\overline{\gamma}_b = 20$ dB, and
$\overline{\gamma}_b/\overline{\gamma}_e = 20$.}\label{fig_side_d}
\end{center}
\end{figure}

In Fig. \ref{fig_side_d}, we plot $T_s^{\ast}$ versus $\beta$ for
different values of $N$. For each point of $T_s^{\ast}$, we choose
the $\left(R_b^{\ast},R_e^{\ast}\right)$ pair that maximizes $T_s$ for the
corresponding $\beta$. We first observe that there exists a unique
$\beta^{\ast\circ}$ that maximizes $T_s^{\ast}$. We then observe
that the value of $\beta^{\ast\circ}$ that maximizes $T_s^{\ast}$ is
around $0.9$, which demonstrates that the secrecy throughput is
improved if approximately 10\% of the total transmit power at the
source is allocated to AN signals. We also observe that the maximal
$T_s^{\ast}$, i.e., $T_s^{\ast\circ}$, increases as $N$ increases.
Furthermore, we observe that the value of $\beta^{\ast\circ}$
slightly decreases as $N$ increases, which shows that in order to maintain the optimal secrecy throughput, the power allocated to AN must increase as the source antenna number increases.

\section{Conclusion}\label{sec:conclusion}

In this work we proposed a secure transmission scheme for a
relay wiretap channel, in which an $N$-antenna source transmits both
information signals and AN signals in the presence of multiple
spatially random single-antenna eavesdroppers. Conditioned on the use of a decode-and-forward protocol at the relay, we determined the optimal
parameters
$\left(\beta^{\ast\circ},R_b^{\ast\circ},R_e^{\ast\circ}\right)$
that maximizes the secrecy throughput,
based on our derived expressions for the transmission outage
probability and the secrecy outage probability. In addition, we
examined the impact of $N$ on the secrecy throughput, showing how
the maximal secrecy throughput increases with $N$. The work reported here provides valuable insights into the design of new physical layer security schemes in which the locations of the eavesdroppers are randomly distributed and not known at the source.


\section*{Acknowledgements}
The work of J. Yuan and R. Malaney was funded by the Australian Research Council Discovery Project DP120102607. The work of N. Yang was funded by the Australian Research Council Discovery Project DP150103905, and the Ian Potter Foundation's travel grant 20160251.

\begin{appendices}

\section{Proof of Theorem \ref{t1}}\label{App_t1}

According to \eqref{snr_e}, \eqref{cdf_3} and \eqref{cdf_4}, we
re-express \eqref{p_sec} as
\begin{align}\label{p_sec_2}
P_{so}&=1-{\Pr}\left\{\Gamma_{E}\leq\tau_e\right\}\notag\\
&=1-{\Pr}\left\{\max_{i\in\Phi} \left\{\max\left\{\gamma_{si},\gamma_{ri}\right\}\right\}\leq\tau_e\right\}\notag\\
&=1-\mathbb{E}_{\Phi}\left[\prod_{i\in\Phi}{\Pr}\left\{\max\left\{\gamma_{si},\gamma_{ri}\right\}\leq\tau_e\right\}\right]\notag\\
&=1-\mathbb{E}_{\Phi}\left[\prod_{i\in\Phi }\left(1-\left(1+\frac{\left(1-\beta\right)\tau_e}{\beta\left(N-1\right)}\right)^{-\left(N-1\right)}\right.\right.\notag\\
&~~\left.\left.\times\exp\left(-\frac{\tau_e}{\beta\overline{\gamma}_{si}}\right)\right)\left(1-\exp\left(-\frac{\tau_e}{\overline{\gamma}_{ri}}\right)\right)\right]\notag\\
&\overset{(a)}{=}1-\exp\left(-2\lambda\left(\mathcal{J}_1+\mathcal{J}_2-\mathcal{J}_3\right)\right).
\end{align}
where
\begin{align}
\label{J_1} \mathcal{J}_1 = &\int_0^{\infty}\int_{0}^{\pi}d_{si}\left(\left(1+\frac{\left(1-\beta\right)\tau_e}{\beta\left(N-1\right)}\right)^{-\left(N-1\right)}\right.\notag\\
&\left.\hspace{2cm}\times
\exp\left(-\frac{\tau_e\sigma_{i1}^2}{\beta P_s}d_{si}^{\eta}\right)
\right)dd_{si}d\theta,
\end{align}
\begin{align}
\label{J_2} \mathcal{J}_2 =
\int_0^{\infty}\int_{0}^{\pi}d_{si}\exp\left(-\frac{\tau_e\sigma_{i2}^2}{P_r}d_{ri}^{\eta}\right)dd_{si}d\theta,
\end{align}
and
\begin{align}
\label{J_3} \mathcal{J}_3 = & \int_0^{\infty}\int_{0}^{\pi}d_{si}\left(1+\frac{\left(1-\beta\right)\tau_e}{\beta\left(N-1\right)}\right)^{-\left(N-1\right)}\notag\\
&\hspace{1 cm}\times\exp\left(-\frac{\tau_e\sigma_{i1}^2}{\beta
P_s}d_{si}^{\eta}-\frac{\tau_e\sigma_{i2}^2}{P_r}d_{ri}^{\eta}\right)dd_{si}d\theta,
\end{align}
and $(a)$ follows by applying the probability generating functional
(PGFL) for the PPP $\Phi$, given by \cite{stoyan}
\begin{align}\label{pgfl}
\mathbb{E}_{\Phi}\left[\prod_{x\in\Phi}f\left(x\right)\right]=\exp\left\{-\int_{\mathbb{R}^2}\left[1-f\left(x\right)\right]\lambda
dx\right\},
\end{align}
and by changing to polar coordinates.

To proceed, we first derive $\mathcal{J}_1$ as
\begin{align}\label{J_1_2}
\mathcal{J}_1&=\pi\left(1+\frac{\left(1-\beta\right)\tau_e}{\beta\left(N-1\right)}\right)^{-\left(N-1\right)}\notag\\
&\hspace{2cm}\times\int_{0}^{\infty}d_{si}\exp\left(-\frac{\tau_e\sigma_{i1}^2}{\beta P_s}d_{si}^{\eta}\right)dd_{si}\notag\\
&\overset{(b)}{=}\frac{\pi}{2}\left(1+\frac{\left(1-\beta\right)\tau_e}{\beta\left(N-1\right)}\right)^{-\left(N-1\right)}\notag\\
&\hspace{2cm}\times\int_{0}^{\infty}\exp\left(-\frac{\tau_e\sigma_{i1}^2}{\beta P_s}u^{\frac{\eta}{2}}\right)du\notag\\
&\overset{(c)}{=}\frac{\pi}{\eta}\left(\frac{\beta P_s}{\tau_e\sigma_{i1}^2}\right)^{\frac{2}{\eta}}\left(1+\frac{\left(1-\beta\right)\tau_e}{\beta\left(N-1\right)}\right)^{-\left(N-1\right)}\notag\\
&\hspace{2cm}\times\int_0^{\infty}\exp\left(-t\right)t^{\frac{2}{\eta}-1}dt\notag\\
&\overset{(d)}{=}\frac{\pi}{\eta}\left(\frac{\beta
P_s}{\tau_e\sigma_{i1}^2}\right)^{\frac{2}{\eta}}
\left(1+\frac{\left(1-\beta\right)\tau_e}{\beta\left(N-1\right)}\right)^{-\left(N-1\right)}\Gamma\left(\frac{2}{\eta}\right),
\end{align}
where in $(b)$ we have used $u=d_{si}^2$, in $(c)$ we have used
$t=\frac{\tau_e\sigma_{i1}^2}{\beta P_s}u^{\frac{\eta}{2}}$, and
$(d)$ follows from the definition of the gamma function. We then
derive $\mathcal{J}_2$ as
\begin{align}\label{J_2_2}
&\mathcal{J}_2\notag\\
=&\int_{0}^{\infty}\!\!\!\!\int_{0}^{\pi}\!\!\!d_{si}\exp\!\left(\!\!-\frac{\tau_e\sigma_{i2}^2}{P_r}\!\left(d_{sr}^2+d_{si}^2\!-\!2d_{sr}d_{si}\cos\theta\right)^{\frac{\eta}{2}}\!\right)\!dd_{si}d\theta.
\end{align}
We further derive $\mathcal{J}_3$ as
\begin{align}\label{J_3_2}
&\mathcal{J}_3\notag\\
=&\left(1+\frac{\left(1-\beta\right)\tau_e}{\beta\left(N-1\right)}\right)^{-\left(N-1\right)}\int_{0}^{\infty}\int_{0}^{\pi}d_{si}\exp\left(-\frac{\tau_e\sigma_{i1}^2}{\beta P_s}d_{si}^{\eta}\right)\notag\\
&\times\exp\!\left(\!-\frac{\tau_e\sigma_{i2}^2}{P_r}\left(d_{sr}^2\!+\!d_{si}^2\!-\!2d_{sr}d_{si}\cos\theta\right)^{\frac{\eta}{2}}\!\right)\!dd_{si}d\theta.
\end{align}

Substituting \eqref{J_1_2}, \eqref{J_2_2}, and \eqref{J_3_2} into
\eqref{p_sec_2}, we obtain the desired result in \eqref{t1_result},
which completes the proof.
\end{appendices}

\end{document}